\newcommand{\diracslash}[1]{#1\llap{/\kern2pt}}

\newcommand{\be}{\begin{equation}}
	\newcommand{\ee}{\end{equation}}
\newcommand{\bea}{\begin{eqnarray}}
	\newcommand{\eea}{\end{eqnarray}}
\newcommand{\ba}[1]{\begin{array}{#1}}
	\newcommand{\ea}{\end{array}}

\documentclass[prd,11pt,aps,floats,nofootinbib,tightenlines,showpacs]{revtex4-1}
\usepackage{epsfig,graphicx,pstricks}
\usepackage{psfrag}
\usepackage{color}
\usepackage{amsmath}
\usepackage{amsfonts}
\usepackage{amssymb}
\usepackage{textcomp}
\usepackage{multirow}
\usepackage{subfigure}
\usepackage[breaklinks=true,colorlinks=true,linkcolor=blue,urlcolor=blue,citecolor=blue]{hyperref}
\usepackage[normalem]{ulem}
\newcommand{\SU}{\mathrm{SU}}
\newcommand{\Tc}{T_{\mbox{\tiny{c}}}}
\newcommand{\hc}{h_{\mbox{\tiny{c}}}}
\newcommand{\kB}{k_{\mbox{\tiny{B}}}}
\newcommand{\THag}{T_{\mbox{\tiny{H}}}}
\newcommand{\rhoHag}{\rho_{\mbox{\tiny{H}}}}
\newcommand{\ZHag}{\mathcal{Z}_{\mbox{\tiny{H}}}}

\newcommand{\mcomment}[1]{}
\begin {document}

\begin{titlepage}

\begin{center}
{\Large\bf Critical exponents and transport properties near the QCD critical endpoint from the statistical bootstrap model}
\end{center}
\vskip1.3cm
\centerline{Guruprasad~Kadam,$^{a,b}$\footnote{\href{mailto:guruprasadkadam18@gmail.com}{{\tt guruprasadkadam18@gmail.com}}} Hiranmaya Mishra,$^{c}$\footnote{\href{mailto:hm@prl.res.in}{{\tt hm@prl.res.in}}} and Marco~Panero$^{d,e}$\footnote{\href{mailto:marco.panero@unito.it}{{\tt marco.panero@unito.it}}}}
\vskip1.5cm
\centerline{\sl $^a$Department of Physics, Shivaji University}
\centerline{\sl Kolhapur, Maharashtra - 416 004, India}
\vskip0.5cm
\centerline{\sl $^b$School of Physical Sciences, National Institute of Science Education and Research Bhubaneswar}
\centerline{\sl HBNI, Jatni 752050, Odisha, India}
\vskip0.5cm
\centerline{\sl $^c$Theory Division, Physical Research Laboratory}
\centerline{\sl Navarangpura, Ahmedabad - 380 009, India}
\vskip0.5cm
\centerline{\sl $^d$Department of Physics, University of Turin and $^e$INFN, Turin}
\centerline{\sl Via Pietro Giuria 1, I-10125 Turin, Italy}
\vskip1.0cm

\setcounter{footnote}{0}

\begin{abstract}
\noindent We present an estimate of the behavior of the shear and bulk viscosity coefficients when the QCD critical point is approached from the hadronic side, describing hadronic matter within the statistical bootstrap model of strong interactions. The bootstrap model shows critical behavior near the quark-hadron transition temperature if the parameter characterizing the degeneracy of Hagedorn states is properly chosen. We calculate the critical exponents and amplitudes of relevant thermodynamic quantities near the QCD critical point and combine them with an \emph{Ansatz} for the shear and bulk viscosity coefficients to derive the behavior of these coefficients near the critical point. The shear viscosity to entropy density ratio is found to decrease when the temperature is increased, and to approach the Kovtun-Son-Starinets bound $1/(4\pi)$ faster near the critical point, while the bulk viscosity coefficient is found to rise very rapidly.
\end{abstract}

\end{titlepage}

\section{Introduction}
\label{sec:introduction}

The phase diagram of strongly interacting matter has attracted a lot of attention in the past couple of decades~\cite{Fukushima:2010bq, Hands:2001ve, Shuryak:1996pb}. At low temperature $T$ and low baryon chemical potential $\mu_B$, quantum chromodynamics (QCD) matter consists of colorless hadrons, while at high temperature and baryon density the degrees of freedom are colored quarks and gluons. Lattice QCD simulations indicate that, at $\mu_B\sim 0$, the transition from the hadronic phase to the quark-gluon plasma phase (QGP) is actually an analytic cross over~\cite{Aoki:2006br, Borsanyi:2010bp, Borsanyi:2015waa, Petreczky:2012fsa, Ding:2015ona, Friman:2011zz}. At finite baryon densities $\mu_B/T \gtrsim 1$, however, lattice QCD calculations are affected by the sign problem~\cite{Hands:2001jn,Aoki:2005dt,Alford:2002ng,deForcrand:2010ys}, and, even though very recent works, based on the analytical continuation from imaginary to real $\mu_B$, are now probing values of the chemical potential up to real $\mu_B \sim 300$~MeV~\cite{Borsanyi:2020fev}, results at large values of $\mu_B$ are still scarce. Hence, in this regime of the phase diagram one has to resort to effective models of QCD, such as the Nambu-Jona-Lasinio model~\cite{Klevansky:1992qe, Hatsuda:1994pi}, the quark-meson-coupling model~\cite{Schaefer:2007pw} etc. Finally, at low temperature and sufficiently large baryon density many phenomenological models predict the transition between the hadronic phase and the deconfined phase to be of first order~\cite{Asakawa:1989bq,Barducci:1989eu,Barducci:1993bh,Berges:1998rc,Halasz:1998qr,Scavenius:2000qd,Antoniou:2002xq,Hatta:2002sj,Bhattacharyya:2010wp}: see also the discussion in ref.~\cite{Bzdak:2019pkr}.

Apart from the effective QCD models mentioned above, a simple model to describe the hadronic phase of QCD is the hadron resonance gas model (HRG). This model is based on the $S$-matrix formulation of statistical mechanics~\cite{Dashen:1969ep}. At low density, as it turns out, the thermodynamics can be approximately modeled in terms of a non-interacting gas of hadrons and resonances~\cite{Dashen:1974yy,Welke:1990za,Venugopalan:1992hy}. The predictions of this model have been compared with lattice QCD simulations, finding good agreement for temperatures up to $T\sim 150$~MeV except for some discrepancies in the trace anomaly~\cite{Karsch:2003vd,Vovchenko:2016rkn,Dash:2018can}. Later studies found that the agreement can be improved, if the contribution of a continuous density of states is included in the mass spectrum of the HRG~\cite{NoronhaHostler:2008ju,NoronhaHostler:2012ug,Kadam:2014cua,Vovchenko:2014pka}. Remarkably, analogous results have been obtained also in lattice simulations of $\SU(N)$ Yang-Mills theories without dynamical quarks~\cite{Meyer:2009tq, Caselle:2015tza}, and even in three spacetime dimensions~\cite{Caselle:2011fy}.

A description of the density of hadron states in terms of a continuous distribution is the basis of the statistical bootstrap model (SBM)~\cite{Hagedorn:1965st,Frautschi:1971ij}, which attracted a lot of attention in the particle physics community in the pre-QCD era. The mass spectrum of abundant formation of heavy resonances and higher angular momentum states can be consistently described by a self-similar structure of hadrons through the bootstrap condition. These high-mass resonances have an interesting effect on the strong interaction thermodynamics: in the thermodynamic system dominated by exponentially rising resonance states there is a finite limiting temperature $\THag$, called Hagedorn temperature. The existence of this limiting temperature indicates that the hadron resonance gas cannot exist at physical temperatures $T>\THag$, and suggests that strongly interacting matter should then enter a different phase. The bootstrap condition of the SBM requires the density of states to be of the form $ \rho(m)\sim m^{a}\: \exp(m/\THag)$~\cite{Satz:1978us, Huang:1970iq, Blanchard:2004du}, where $a$ is a constant. Interestingly, the string model (or dual resonance model) of strong interactions~\cite{Mandelstam:1974fq} also predicts this type of density of states. The $a$ constant plays an important role in determining the thermodynamics of the SBM near the Hagedorn temperature. In fact, for the choice $a=-4$ both the energy density and the entropy density remain finite near $\THag$ and one expects a phase transition to take place~\cite{Antoniou:2002xq, Satz:1978us, Castorina:2009de}, so that $\THag$ can be interpreted as a critical temperature, $\Tc$.

A particularly interesting point in the QCD phase diagram is the conjectured critical end point (CEP). It should be remarked that, so far, the existence of the CEP has neither been proven theoretically, nor has it been observed experimentally. However, its existence is strongly suggested by the aforementioned model calculations investigating the phase diagram region at low temperatures and baryon densities larger than that of nuclear matter, which predict a first-order transition line separating the hadronic phase from a deconfined phase: since that line is known not to extend all the way to the $\mu_B=0$ axis (where the transition is actually a crossover), it should end at a CEP, where the transition should be a continuous one~\cite{Stephanov:1998dy}. A lot of theoretical investigation has been carried out, and is still going on, to locate the CEP and predict possible experimental signatures, see refs.~\cite{Gavai:2016lwu,Mohanty:2009vb,Stephanov:2004xs,Luo:2017faz} for reviews. On the experimental side, an entire experimental program, namely the Beam Energy Scan (BES) program, has been devoted at the Super Proton Synchrotron (SPS) and at the Relativistic Heavy Ion Collider (RHIC) to search for the CEP~\cite{Aggarwal:2010cw, Heinz:2015tua}. In particular, as suggested in ref.~\cite{Stephanov:1998dy}, the existence and the location of the CEP could be revealed by the observation of a suppression in temperature and chemical potential fluctuations on an event-by-event basis, and by large fluctuations in the multiplicity of low-energy pions.

A very important feature of the critical point is the emergence of a universal critical mode. As the system approaches the critical point, this mode rises very rapidly with some power of the correlation length $\xi$, which eventually diverges at the critical point. For instance, the variance, skewness and kurtosis of the non-Gaussian fluctuation of the critical mode grow as $\xi^{2}$, $\xi^{9/2}$ and $\xi^{7}$, respectively~\cite{Stephanov:2008qz,Asakawa:2015ybt,Bluhm:2020mpc}. In the experimental search for the critical endpoint, these critical fluctuations can be accessed by measuring event-by-event fluctuations of particle multiplicities~\cite{Heinz:2015tua}.

While ``static'' critical phenomena have been extensively studied theoretically, an avenue that has been explored less is the one of ``dynamical'' critical phenomena. As it turns out, critical singularities can also occur in quantities encoding the dynamical properties of the medium, like transport coefficients. Away from the critical point, the dynamic properties of a system can be characterized by hydrodynamics, which provides an effective description of the fluid in the low-frequency, long-wavelength limit. Hydrodynamics describes fluctuations of conserved quantities at equilibrium and any additional slow variable that occurs due to the existence of a spontaneously broken symmetry. In the hydrodynamic effective theory the dynamical critical fluctuations are described by coupling the order-parameter field with the conserved momentum density. In this model, which is called the H model in the classification of dynamical critical phenomena~\cite{Hohenberg:1977ym} by Hohenberg and Halperin, the transport coefficients depend on the correlation length as
\begin{equation}
\eta \sim \xi^{\frac{\epsilon}{19}},\hspace{0.5cm} \kappa \sim \xi,\hspace{0.5cm} D_{B}\sim \frac{1}{\xi}, \hspace{0.5cm}\zeta\sim \xi^{3}.
\end{equation}
This behavior suggests that the transport coefficients would affect the bulk hydrodynamic evolution of the matter created in heavy-ion collisions near the QCD critical point~\cite{Monnai:2016kud, Monnai:2017ber,Paech:2003fe,Nonaka:2004pg,Paech:2005cx,Herold:2013bi,Nahrgang:2016ayr,Kapusta:2012zb}.

It is worth emphasizing that, while lattice calculations remain \emph{the} tool of choice for theoretical first-principle studies of various quantities relevant for strong-interaction matter, their applicability in studies of transport coefficients in the proximity of the QCD critical endpoint is severely limited, for a two-fold reason. On one hand, as we remarked above, the existence of the sign problem poses a formidable barrier to lattice simulations at finite baryon-number density: a barrier that might even be impossible to overcome with classical computers, if it is related to fundamental computational-complexity issues~\cite{Troyer:2004ge}. On the other hand, even at zero baryon-number density, the lattice determination of transport coefficients of QCD matter involves its own difficulties, due to the fact that lattice QCD calculations are done in a Euclidean spacetime, and typically the extraction of quantities involved in real-time dynamics requires a Wick rotation back to Minkowski signature, with the reconstruction of a continuous spectral function from a finite set of Euclidean data, which is an ill-posed numerical problem~\cite{Meyer:2011gj}. Despite some recent progress (see, e.g., refs.~\cite{Cuniberti:2001hm, Meyer:2007ic, Meyer:2007dy, Burnier:2011jq, Panero:2013pla, Rothkopf:2016luz}), a general solution to this type of problems is still unknown.

For these reasons, phenomenological models remain a useful theoretical tool to get some insight into the physics near the QCD critical endpoint. In this work, we extract critical exponents~\cite{Satz:1978us} and amplitudes of thermodynamic quantities relevant near the critical point within the statistical bootstrap model. We then derive the singularities characterizing shear and bulk viscosity coefficients, starting from an \emph{Ansatz} for viscosity coefficients~\cite{Antoniou:2016ikh} that is suitable for a hydrodynamic system with conserved baryon charge. We then estimate viscosity coefficients  near the critical point from the hadronic side using the critical exponents of this model.

We organize the paper as follows. In section~\ref{sec:SBM} we review the derivation of the critical exponents (and amplitudes) close to the critical point in the critical bootstrap model, that was first worked out in ref.~\cite{Satz:1978us}, with a few additional remarks and comments. In section~\ref{sec:transport_coefficients} we derive the singularities of shear and bulk viscosity near the critical point. In section~\ref{sec:results} we present our results. Finally, in section~\ref{sec:discussion_and_conclusions} we summarize our findings and conclude. Throughout the paper, we work in natural units ($\hbar=c=\kB=1$).
 
\section{Statistical bootstrap model: criticality and critical point exponents}
\label{sec:SBM}
\subsection{Critical exponents}

The analysis of critical phenomena is based on the assumption that, in the $T\rightarrow \Tc$ limit, any relevant thermodynamic quantity can be separated into a regular part and a singular part. The singular part may be divergent or it can have a divergent derivative. It is further assumed that the singular part of all the relevant thermodynamic quantities is proportional to some power of $t$, where $t=(T-\Tc)/\Tc$. These powers, called critical exponents, characterize the nature of singularity at the critical point. The critical exponents, $\hat{\alpha}$, $\hat{\beta}$, $\hat{\gamma}$ and $\hat{\nu}$ are defined through the following power laws~\cite{Huang:1987sm} (in the limit $t\rightarrow 0^{-}$):
\begin{eqnarray}
\label{alpha_exponent}
C_V &=&\mathcal{C_{-}} \: |t|^{-\hat{\alpha}}, \\
\label{beta_exponent}
1-\frac{n_{B}}{n_{B,c}} &=&\mathcal{N_-} \: |t|^{\hat{\beta}}, \\ 
\label{gamma_exponent}
k_{T} &=&\mathcal{K}_{-} \: |t|^{-\hat{\gamma}}, \\
\label{nu_exponent}
\xi &=&\Xi_{-}\:|t|^{-\hat{\nu}},
\end{eqnarray}
where $C_V$, $n_{B,c}$, $k_{T}$ and $\xi$ respectively denote the specific heat, the critical baryon density, the isothermal compressibility and the correlation length, while $\mathcal{C_{-}}$, $\mathcal{N_-}$, $\mathcal{K}_{-}$ and $\Xi_{-}$ are the corresponding amplitudes from the hadronic side ($T <\Tc$). Note that eq.~(\ref{beta_exponent}) is an equation of state, relating baryon density $n_{B}$ and pressure $p$ near the critical point.

\subsection{Formulation of the model}
\label{subsec:formulation_of_the_model}

We follow ref.~\cite{Satz:1978us} to extract the amplitudes and critical exponents within the statistical bootstrap model. We first discuss the case of vanishing baryonic chemical potential, $\mu_B=0$. Consider an ideal gas of hadrons and all possible resonance states as non-interacting constituents: the partition function of this system can be written as~\cite{Beth:1937zz}
\begin{equation}
\mathcal{Z}(T,V)=\sum_{N=1}^\infty \frac{V^N}{N!}\prod_{i=1}^{N}\int \frac{d^3p_i}{(2\pi)^3}\: dm_{i}\:\rho_{i}(m_{i}) e^{-E_{i}/T}
\label{pf}
\end{equation}
where $\rho(m)$ is the hadron spectrum included in the HRG model. In the simplest formulation of the model, that was discussed in ref.~\cite{Satz:1978us}, only pions were considered as the ``basic'' hadrons. More recently, however, it has become customary to include all the hadrons and resonances that have been detected experimentally up to some energy scale $M$ and take $\rho(m)=\sum_{j}\delta(m-m_j)$. Such discrete mass spectrum leads to the physical hadron resonance gas model. In the physical HRG, if $g_i$ is the degeneracy of the $i$-th hadronic species, then for spin degrees of freedom the degeneracy factors turn out to be $g_i \sim m_{i}^2$~\cite{Castorina:2009de}. Thus, one sees that the spin multiplicity already can result in an unbounded increase in resonances. The upshot of the $m^2$ dependence of resonance degeneracy is that the partition function of the physical resonance gas and all of its higher-order derivatives remain finite at $\Tc$. Thus, the required degeneracy structure is absent in the physical resonance gas and hence it does not show critical behavior.

It turns out that the degeneracy structure required to show critical behavior is present in the Hagedorn density of states, which can be used to model the spectral density above $M$ in terms of a continuous distribution. Consider a density of states of the form
\begin{equation}
\label{states}
\rho(m)= \sum_{i} \left[ g_{i} \cdot \delta(m-m_{i})\right] +\theta(m-M)\rhoHag(m),
\end{equation}
where the sum ranges over all hadrons species with mass $m_i \le M$, $g_i$ denotes the degeneracy of each species, while $\rhoHag$ is the continuous contribution to the density of states. For our analysis, we included all hadronic states reported in ref.~\cite{Zyla:2020ssz} with masses not larger than $M=2.25$~GeV. It should be noted that choosing a different $M$ value in the same ballpark would not lead to significant differences e.g. in the equilibrium thermodynamic quantities in the low-temperature phase. The reason for this is that at low temperatures the thermodynamics is dominated by the lightest hadrons, and including or not including the contribution from some discrete heavy states does not have significant impact on the equation of state at low $T$.

Note that, in the simplest possible formulation of the model, the discrete part of the spectrum could include only pions, and one could model all the states of the spectrum above the two-pion threshold (setting $M=2m_\pi$) in terms of a Hagedorn density of states:
\begin{equation}
\label{pion_states}
\rho_{\mbox{\tiny{simplest}}}(m)=g_{\pi} \cdot \delta(m-m_{\pi})+\theta(m-2m_\pi)\rhoHag(m)
\end{equation}
where $g_\pi$ denotes the pion degeneracy, which is equal to $3$. While such picture is clearly a very crude model of the hadron spectrum, it still captures some interesting finite-temperature features, at least at a qualitative level, and is useful to highlight some general consequences for the thermodynamics and transport properties.

The logarithm of the partition function~(\ref{pf}) with the density of states in eq.~(\ref{states}) is written as
\begin{equation}
\label{ln_Z_as_sum}
\ln\mathcal{Z}(T,V)=\ln\mathcal{Z}_{\mbox{\tiny{discrete}}}(T,V)+\ln\ZHag(T,V)
\end{equation}
in which the first summand on the right-hand side, which does not depend on the \emph{Ansatz} for the continuous part of the density of states, encodes the contribution from a gas of non-interacting hadrons in the discrete part of the spectrum (i.e. hadrons whose masses are not larger than $M$). In particular, the contribution to $\ln\mathcal{Z}_{\mbox{\tiny{discrete}}}$ due to pions can be written in the form
\begin{equation}
\label{pion_contribution_to_ln_Z}
\ln\mathcal{Z}_{\pi}(T,V)=-g_{\pi}\int \frac{d^3p}{(2\pi)^{3}}\ln\left[1-\exp \left( \frac{\sqrt{p^2+m_\pi^2}}{T} \right) \right]=\frac{g_{\pi}m_\pi^2 TV}{2\pi^2}\sum_{n=1}^\infty \frac{K_{2}(n m_\pi /T)}{n^2},
\end{equation}
where $K_{n}(z)$ is a modified Bessel function of the second kind of order $n$. For large real values of its argument, one has
\begin{equation}
\label{KN_at_large_argument}
K_{n}(z) = \sqrt{\frac{\pi}{2z}}\:e^{-z}\left[1+\mathcal{O}(z^{-1})\right] .
\end{equation}
The contributions to $\ln\mathcal{Z}_{\mbox{\tiny{discrete}}}$ from the other hadron species in the discrete part of the spectrum can be derived in a similar way, and one obtains
\begin{equation}
\label{discrete_contribution_to_ln_Z}
\ln\mathcal{Z}_{\mbox{\tiny{discrete}}}(T,V)=\sum_i \frac{g_{i}m_i^2 TV}{2\pi^2}\sum_{n=1}^\infty (-\eta_i)^{n+1}\frac{K_{2}(n m_\pi /T)}{n^2},
\end{equation}
where the sum over $i$ ranges over all hadrons with mass $m_i \le M$, as in eq.~(\ref{states}), and $\eta_i=-1$ for bosons, while $\eta_i=1$ for fermions.

The second summand appearing on the right-hand side of eq.~(\ref{ln_Z_as_sum}) represents the contribution due to the continuous part of the spectrum:
\begin{equation}
\ln\ZHag(T,V)=V\Phi_{B}(T)
\label{pf1}
\end{equation}
with
\begin{equation}
\Phi_{B}(T)=\int_{M}^{\infty}dm \rhoHag(m)\int \frac{d^{3}p}{(2\pi)^3}e^{-E_{i}/T}
\end{equation}
where, as above, $M$ is the threshold separating the discrete (for $m\le M$) and the continuous (for $m > M$) parts of the spectrum. Performing the momentum integration, one obtains
\begin{equation}
\Phi_{B}(T)=\frac{T}{2\pi^2}\int_{M}^{\infty}dm\:m^2 \rhoHag(m) K_{2}(m/T).
\end{equation}
Using eq.~(\ref{KN_at_large_argument}), for $m/T\gg1$ one gets
\begin{equation}
\Phi_{B}(T)=\left(\frac{T}{2\pi}\right)^{3/2}\int_{M}^{\infty}dm\:m^{3/2} \rhoHag(m) e^{-m/T}.
\end{equation}
All the thermodynamic functions can be readily obtained from the partition function in eq.~(\ref{pf1}) once the continuous part of the mass spectrum $\rho(m)$ is specified.

In the statistical bootstrap model (see refs.~\cite{Tawfik:2014eba, Rafelski:2015cxa} for reviews), hadronic matter at high temperature is dominated by formation of resonances whose number grows exponentially. The bootstrap condition leads to a solution for the mass spectrum of the form~\cite{Hagedorn:1965st, Frautschi:1971ij}
\begin{equation}
\rhoHag(m)=A\:m^{a}\:e^{bm}
\label{hagspec}
\end{equation}
where $A$, $a$, and $b$ are constant parameters. In particular, the parameter $A$ provides the normalization of the resonance contributions relative to that of the pions. The parameter $a$ specifies the nature of the degeneracy of high-mass resonances, and also determines the critical behavior of hadronic matter. One possible solution of the bootstrap condition was derived in ref.~\cite{Nahm:1972zc}, yielding $a \simeq 3$. Finally, the parameter $b$ turns out to be the inverse of the Hagedorn temperature at which thermodynamic functions show singular behavior.

Restoring the dependence on the fugacity $z_B=\exp(\mu_B/T)$, the contribution to the partition function associated with the continuous spectrum~(\ref{hagspec}) can be written as
\begin{equation}
\ln\ZHag(T,V,z_B)=AVz_B\left(\frac{T}{2\pi}\right)^{3/2}\int_{M}^{\infty}dm\:m^{a+3/2}  e^{(b-1/T) m}.
\label{pf4}
\end{equation}
At this point, we should stress an important observation: in order to obtain eq.~(\ref{pf4}), in which $z_B$ is factorized on the right-hand side, it has been implicitly assumed that $b$ is \emph{independent} from $\mu_B$, i.e. that the critical temperature $\Tc$ does not depend on the fugacity $z_B$. Strictly speaking, however, this is not fully justified: as has been discussed in detail in the literature~\cite{Kapoyannis:1997qj,Kapoyannis:1998py,Kapoyannis:1999ar}, in the presence of arbitrary fugacity $z_B$, the bootstrap equation takes the form
\begin{equation}
\label{generalized_bootstrap_equation}
\phi(T,z_B)=2G(T,z_B)-\exp[G(T,z_B)]+1,
\end{equation}
where $\phi$ is an {\it{input function}}, receiving contributions from the physical hadrons of the theory, while $G$, which encodes their interactions in terms of the bootstrap picture (whereby strongly interacting systems of particles form larger clusters of particles, which in turn form larger clusters, etc.) is the Laplace transform of the mass spectrum. Eq.~(\ref{generalized_bootstrap_equation}) has a square-root branch point singularity for $\phi=2\ln 2-1$ (or, equivalently, for $G=\ln2$), which defines the boundary of the hadronic phase in this model through a non-trivial relation between $T$ and $z_B$. In other words, strictly speaking, the critical temperature $T_c$ is a non-trivial function of $z_B$. In eq.~(\ref{pf4}) and in the rest of this work, however, we assume that the dependence of $T_c$ on the fugacity is mild, i.e. we work in the approximation in which $b=1/T_c$ is constant. While this simplification may appear to be crude, it is worth noting that during the past few years lattice QCD calculations have conclusively proven that the change of state between the hadronic, broken-chiral-symmetry phase and the deconfined, chirally symmetric phase at zero chemical potential is a crossover~\cite{Bernard:2004je,Cheng:2006qk,Aoki:2006we}, and that at small but finite values of $\mu_B$ the curvature of the line describing the crossover in the QCD phase diagram is very small~\cite{Kaczmarek:2011zz, Endrodi:2011gv, Cea:2014xva, Bonati:2014rfa, HotQCD:2018pds}. As a consequence, it is not unreasonable to expect that, even within the approximation of a critical temperature independent from the chemical potential, the statistical bootstrap model may still capture the physics close to a possible critical endpoint of QCD at finite chemical potential. Assuming $\Tc$ to be approximately independent from $\mu_B$ simplifies the expression of the partition function, and allows one to get more analytical insight into the physical quantities of interest. In a nutshell, the fact that $z_B$ factors out in the expression of the logarithm of $\ZHag$ implies that the dependence on the chemical potential in this model is somewhat ``trivial''. While the validity of this approximation at large values of $\mu_B$ is not obvious, lattice results lead us to think that its use at least for small and intermediate values of $\mu_B$ should be a reasonable approximation. 

With these caveats in mind, in the next section we shall calculate the critical exponents by taking appropriate derivatives of the partition function~(\ref{pf4}) and then taking the $T\to 1/b$ limit.

\subsection{Critical exponents in the statistical bootstrap model}
\label{subsec:critical_exponents_in_SBM}

With the change of variable $w=m(1/T-b)$, we get
\begin{eqnarray}
\ln\ZHag(T,V,z_B) &=& AVz_B\left(\frac{T}{2\pi}\right)^{3/2} (1/T-b)^{-(a+5/2)}\int_{M(1/T-b)}^{\infty}dw\:w^{a+3/2} \: e^{-w} \nonumber \\
&=& AVz_B\left(\frac{T}{2\pi}\right)^{3/2} (1/T-b)^{-(a+5/2)} \Gamma\left( a+\frac{5}{2} ,M (1/T-b)\right), \label{pf3}
\end{eqnarray}
having expressed the integral in terms of the upper incomplete $\Gamma$ function. The energy density can then be written as
\begin{equation}
\varepsilon=\frac{T^2}{V}\frac{\partial \ln \ZHag}{\partial T}
\end{equation}
and for $T\to 1/b$ one finds that
\begin{equation}
\label{en}
\varepsilon \simeq \left\{
\begin{array}{ll}
\frac{Az_B}{(2\pi b)^{3/2}}\:\Gamma\left(a+\frac{7}{2}\right)\:(1/T-b)^{-(a+\frac{7}{2})}, & \hspace{0.5cm} \mbox{for } a>-7/2\\
-\frac{Az_B}{(2\pi b)^{3/2}}\:\ln[M(1/T-b)], & \hspace{0.5cm}  \mbox{for } a=-7/2\\
\text{constant}, & \hspace{0.5cm} \mbox{for } a<-7/2
\end{array}
\right. .
\end{equation}
Hence for $a<-7/2$ the energy density remains finite (and approaches some critical value $\varepsilon_{\mbox{\tiny{c}}}$) as $T\rightarrow \THag$, implying that the system cannot exist in this state for $\varepsilon>\varepsilon_{\mbox{\tiny{c}}}$ and suggesting that a phase transition must take place.

The specific heat at constant volume can then be written as
\begin{equation}
C_V=\frac{2\varepsilon}{T} +\frac{T^2}{V} \frac{\partial^2}{\partial T^2} \ln \ZHag
\end{equation}
and for $T\to 1/b$ one gets
\begin{equation}
\label{cv}
C_V \simeq \left\{
\begin{array}{ll}
\frac{Ab^2z_B}{(2\pi b)^{3/2}}\:\Gamma\left(a+\frac{9}{2}\right)\:(1/T-b)^{-(a+9/2)}, & \hspace{0.5cm}  \mbox{for } a>-9/2\\
-\frac{A b^2z_B}{(2\pi b)^{3/2}}\ln[M(1/T-b)], & \hspace{0.5cm}  \mbox{for } a=-9/2\\
\text{constant}, & \hspace{0.5cm} \mbox{for } a<-9/2
\end{array}
\right. .
\end{equation}
Comparing eq.~(\ref{cv}) with eq.~(\ref{alpha_exponent}) we deduce the amplitude $\mathcal{C_-}$ as
\begin{equation}
\mathcal{C_-}=\left\{
\begin{array}{ll}
\frac{Ab^2z_B}{(2\pi b)^{3/2}}\:\Gamma\left(a+\frac{9}{2}\right), &\hspace{0.5cm} \text{for} \:a \geqslant-9/2\\
\frac{Ab^2z_B}{(2\pi b)^{3/2}}, & \hspace{0.5cm} \text{for} \:a <-9/2
\end{array}
\right.
\end{equation}
while the critical exponent $\hat{\alpha}$ reads
\begin{equation}
\hat{\alpha}=\left\{
\begin{array}{ll}
a+\frac{9}{2}, &\hspace{0.5cm} \text{for} \:a \geqslant-9/2\\
0, & \hspace{0.5cm} \text{for} \:a <-9/2
\end{array}
\right. .
\end{equation}

The baryon number density $n_B$ can be evaluated as
\begin{equation}
n_{B}=\frac{z_B}{V}\frac{\partial}{\partial z_B} \ln\ZHag =Az_B\left(\frac{T}{2\pi}\right)^{3/2}\:(1/T-b)^{-(a+5/2)}\:\Gamma\left(a+\frac{5}{2},M(1/T-b)\right),
\end{equation}
hence for $T$ close to $1/b$ we get the critical density as
\begin{equation}
n_{B,\mbox{\tiny{c}}} \simeq \left\{
\begin{array}{ll}
\frac{Az_B}{(2\pi b)^{3/2}}\:\Gamma\left(a+\frac{5}{2}\right)\:(1/T-b)^{-(a+5/2)}, & \hspace{0.5cm}  \text{for } a>-5/2\\
-\frac{Az_B}{(2\pi b)^{3/2}}\:\ln[M(1/T-b)], & \hspace{0.5cm} \text{for } a=-5/2\\
\text{constant}, & \hspace{0.5cm} \text{for } a<-5/2
\end{array}
\right. .
\end{equation}
The inverse of the isothermal compressibility is defined as
\begin{equation}
k_{T}^{-1}=-V\:\left(\frac{\partial p}{\partial V}\right)_{T}
\end{equation}
and for a non-interacting resonance gas it takes the following, very simple form:
\begin{equation}
k_{T}^{-1}=n_{B}T.
\end{equation} 
For temperatures close to $1/b$, one obtains,
\begin{equation}
k_{T}^{-1} \simeq \left\{
\begin{array}{ll}
\frac{Az_B}{b(2\pi b)^{3/2}}\:\Gamma\left(a+\frac{5}{2}\right)\:(1/T-b)^{-(a+5/2)}, & \hspace{0.5cm} \text{for }  a>-5/2\\
-\frac{Az_B}{b(2\pi b)^{3/2}}\:\ln[M(1/T-b)], & \hspace{0.5cm}  \text{for } a=-5/2\\
\text{constant}, & \hspace{0.5cm}  \text{for } a<-5/2
\label{kt}
\end{array}
\right. ,
\end{equation}
from which it is straightforward to deduce the amplitude $\mathcal{K_-}$
\begin{equation}
\mathcal{K_-}=\left\{
\begin{array}{ll}
\bigg[\frac{Az_B}{b(2\pi b)^{3/2}}\:\Gamma\left(a+\frac{5}{2}\right)\bigg]^{-1}, &\hspace{0.5cm} \text{for} \:a >-5/2\\
\bigg[\frac{Az_B}{b(2\pi b)^{3/2}}\bigg]^{-1}, & \hspace{0.5cm} \text{for} \:a <-5/2
\end{array}
\right.
\end{equation}
and the critical exponent $\hat{\gamma}$ as
\begin{equation}
\hat{\gamma}=\left\{
\begin{array}{ll}
-\left(a+\frac{5}{2}\right), &\hspace{0.5cm} \text{for} \:a >-5/2\\
0, & \hspace{0.5cm} \text{for} \:a <-5/2
\end{array}
\right. .
\end{equation}

We note that a continuous density of states with $a<-7/2$ makes the energy and entropy densities finite, while all higher-order derivatives diverge near $\THag$. In our analysis of transport coefficients we shall consider the $a=-4$ case~\cite{Antoniou:2002xq,Castorina:2009de,Fiore:1984yu} which leads to normal behavior of the hadronic system near the boundaries of the quark-hadron phase transition line, since it does not allow the energy density to become infinite even for pointlike particles.

At this point, an important observation is in order. Hadrons are \emph{not} elementary, pointlike particles: rather, they arise as color-singlet bound states of the strong interaction, and, for this reason, they can be associated with a characteristic finite size, of the order of the fm. As a consequence of the very nature of hadrons as complex bound states of relativistic, strongly interacting constituents (which defies a description in terms of sufficiently simple phenomenological models), the measurement and even the definition of hadron sizes are, in general, non-trivial (see, for example, ref.~\cite{Pohl:2010zza} for an experimental determination of the radius of a well-known hadron: the proton). It is worth noting that, if corrections related to the finiteness of the particles' physical size are taken into account in our model, the restriction on the admissible values of $a$ become milder, in the sense that finite-particle-size corrections make some of the divergent quantities obtained in the pointlike approximation finite. The fact that finite-particle-size effects can have even a qualitative impact on the details of the description of the thermodynamics of the confining phase of QCD is hardly surprising, as it is well known that they have a significant role in fits of particle multiplicities produced in heavy-ion collisions~\cite{Rischke:1991ke,Yen:1998pa, Yen:1997rv}, and even in the interpretation of non-perturbative theoretical predictions from lattice simulations~\cite{Alba:2016fku}. For this reason, in a more complete discussion, \emph{a priori} one should not discard the $a$ values that lead to unphysical infinities for a system of pointlike particles. However, a fully systematic discussion of finite-particle-size effects would involve a non-trivial amount of additional technicalities (and a certain degree of arbitrariness in the way to define these effects), and lies beyond the scope of our present work. For this reason, in the following we restrict our attention to the simpler, idealized case of pointlike particles, which is nevertheless expected to provide a reasonable approximation of the physics that is studied in currents experiments, especially in view of the fact that the typical sizes of the systems produced in nuclear collisions are significantly larger than hadron sizes~\cite{Kolb:2003dz}, and which does not introduce additional parameters in the description.

\section{Transport coefficients near the critical point}
\label{sec:transport_coefficients}
Approaching the critical point, the thermodynamic quantities relevant for the computation of transport coefficients are: energy density ($\varepsilon$), baryon number density ($n_B$), specific heats ($C_V$ and $C_p$), isothermal compressibility ($k_T$), speed of sound ($C_s$) and correlation length ($\xi$). A set of \emph{Ans\"atze} for the transport coefficients near the critical point can be written in terms of thermodynamic quantities as~\cite{Antoniou:2016ikh} 
\begin{eqnarray}
\label{eta_visc_Tc}
\frac{\eta}{s}&=&\frac{T}{C_s\xi^2s}\mathcal{F_{\eta}}\left(\frac{C_p}{C_V}\right),\\
\label{zeta_visc_Tc}
\frac{\zeta}{s}&=&\frac{hC_s\xi}{s}\mathcal{F_{\zeta}}\left(\frac{C_p}{C_V}\right).
\end{eqnarray}
Near the critical point, the correlation length $\xi$ is the only relevant length scale. Further, longitudinal perturbations can be assumed to be those of the non-equilibrium modes near $\Tc$.  A particularly simple form of the  functions $F_{\eta,\zeta}$, namely $F_{\eta}(C_p/C_V)=f_{\eta}\times (C_p/C_V)$ and $F_{\zeta}=f_{\zeta}\times (C_p/C_V)$, can be obtained from a perturbative treatment of conventional fluids. Here, $f_\eta$ and $f_\zeta$ are non-universal dimensionless constants and depend on the microscopic length scale of the system. Substituting the singular part of the thermodynamic quantities from eqs.~(\ref{alpha_exponent})--(\ref{nu_exponent}) into eqs.~(\ref{eta_visc_Tc}) and~(\ref{zeta_visc_Tc}) we get, as $t \to 0^-$ (i.e $T \to \Tc$ from the hadronic side)
\begin{eqnarray}
\left(\frac{\eta}{s}\right)_{-} &=& \frac{f_{\eta} \mathcal{K}_{-} \lambda_{c}}{\Xi_{-}^2 s_c}\sqrt{\frac{{\Tc}^3 \hc}{\mathcal{C}_{-}} \left(1+\frac{\mathcal{C}_{-}|t|^{\hat{\gamma}-\hat{\alpha}}}{\Tc \lambda_{c}^{2}\mathcal{K}_{-}}\right)^{-1}}\:|t|^{-\hat{\gamma}+2\hat{\nu}+\hat{\alpha}/2}, \label{etatc} \\
\left(\frac{\zeta}{s}\right)_{-} &=& \frac{f_{\zeta} \mathcal{K}_{-} \Xi_{-}\lambda_{c}^3}{s_c}\sqrt{\frac{{\Tc}^3 \hc}{\mathcal{C}_{-}^3} \left(1+\frac{\mathcal{C}_{-}|t|^{\hat{\gamma}-\hat{\alpha}}}{\Tc \lambda_{c}^{2}\mathcal{K}_{-}}\right)^{-3}}\:|t|^{-\hat{\gamma}-\hat{\nu}+3\hat{\alpha}/2}.
\label{zetatc}              
\end{eqnarray}
Here, $\hc$ and $s_c$ respectively denote the enthalpy and entropy densities at $\Tc$, both of which are finite when one sets $a=-4$ in the Hagedorn density of states, while $\lambda_c=(\partial p/\partial T)_{V}$ at $T=\Tc$. The amplitudes $f_\eta$ and $f_\zeta$ are free parameters, which can be fixed by imposing some constraint on the viscosity coefficients near $\Tc$. For instance, as we already mentioned, the gauge-string duality~\cite{Maldacena:1997re,Gubser:1998bc,Witten:1998qj} suggests a universal lower bound $1/(4\pi)$ for the $\eta/s$ ratio~\cite{Kovtun:2004de}. Similar constraints can be imposed on the $\zeta/s$ ratio, too.

\section{Results}
\label{sec:results}
 
\begin{table}[h]
	\begin{center}
		\begin{tabular}{|c|c|c|c|}
			\hline
			density of states & $a$ & $b~[\text{GeV}^{-1}]$ & $A_{1}~[\text{GeV}^{3}]$ \\ \hline
			$\rho_{1}$ & $-4$ & $6.25$ & $0.06144$ \\ \hline 
		\end{tabular}
		\caption{Parameters of the continuous part of the density of states, taken from refs.~\cite{Castorina:2009de, NoronhaHostler:2012ug}. According to the discussion in section~\ref{sec:SBM}, the parameter $b$ is set to the inverse of $\Tc$, whose value is $\Tc=0.160$~GeV. Note that the value of $A_1$ chosen for the $\rho_1$ model corresponds to $A_1=15 \Tc^3$, as discussed in the text.}
		\label{tbl_hag}
	\end{center}
\end{table}
 
\begin{figure}[h!]
\begin{center}
\centerline{\includegraphics[width=0.48\textwidth]{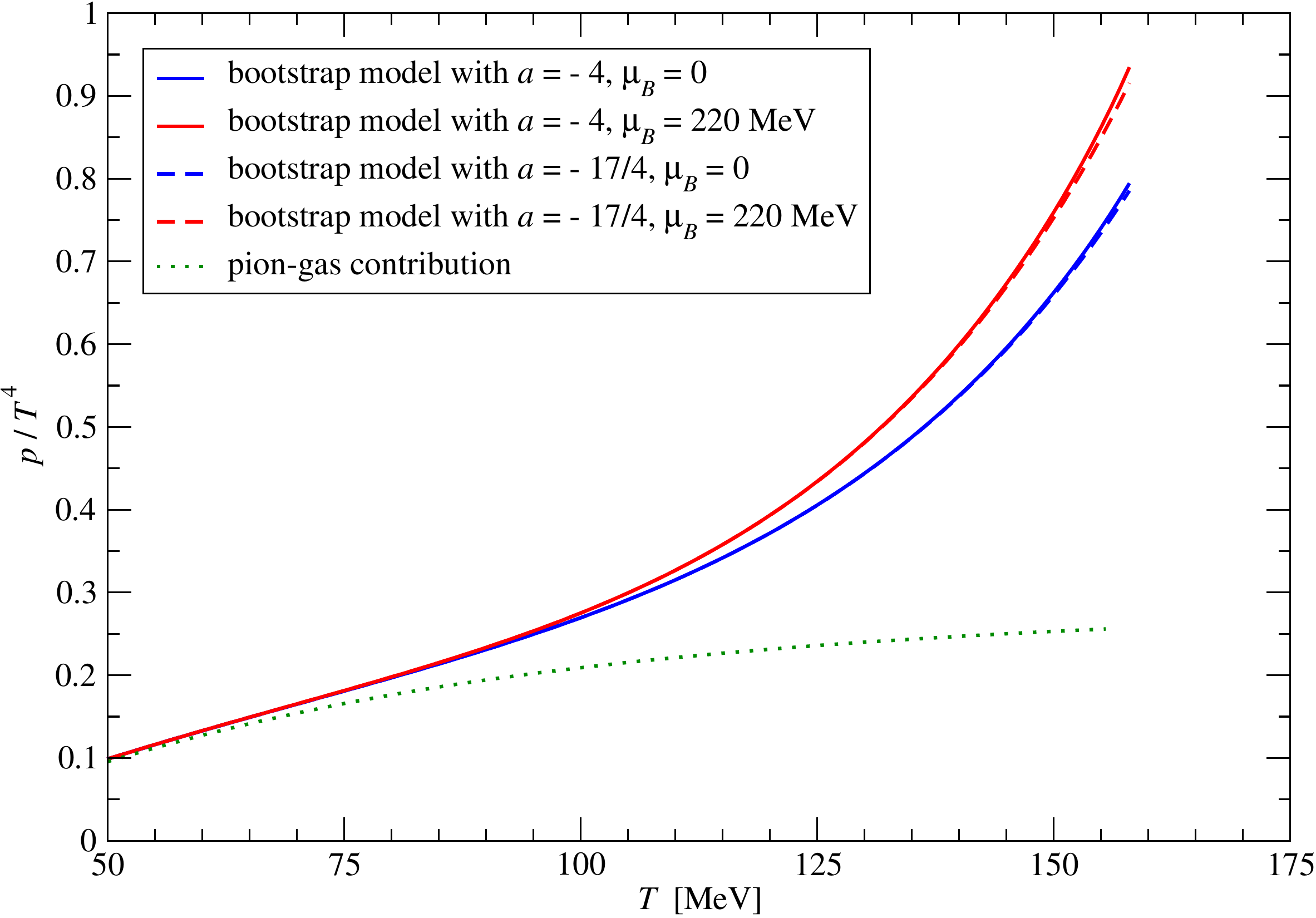} \hfill \includegraphics[width=0.48\textwidth]{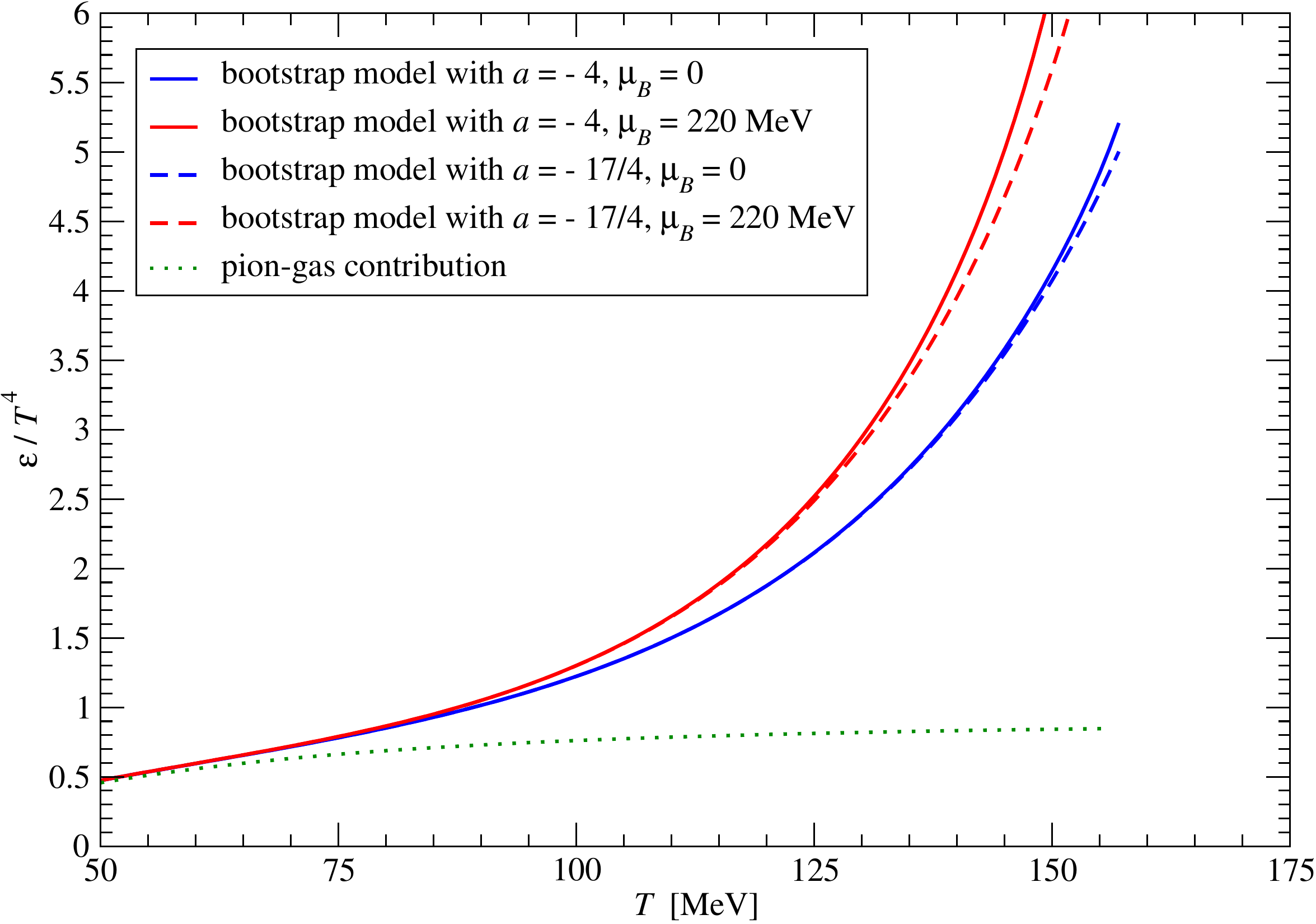}}
\centerline{\includegraphics[width=0.48\textwidth]{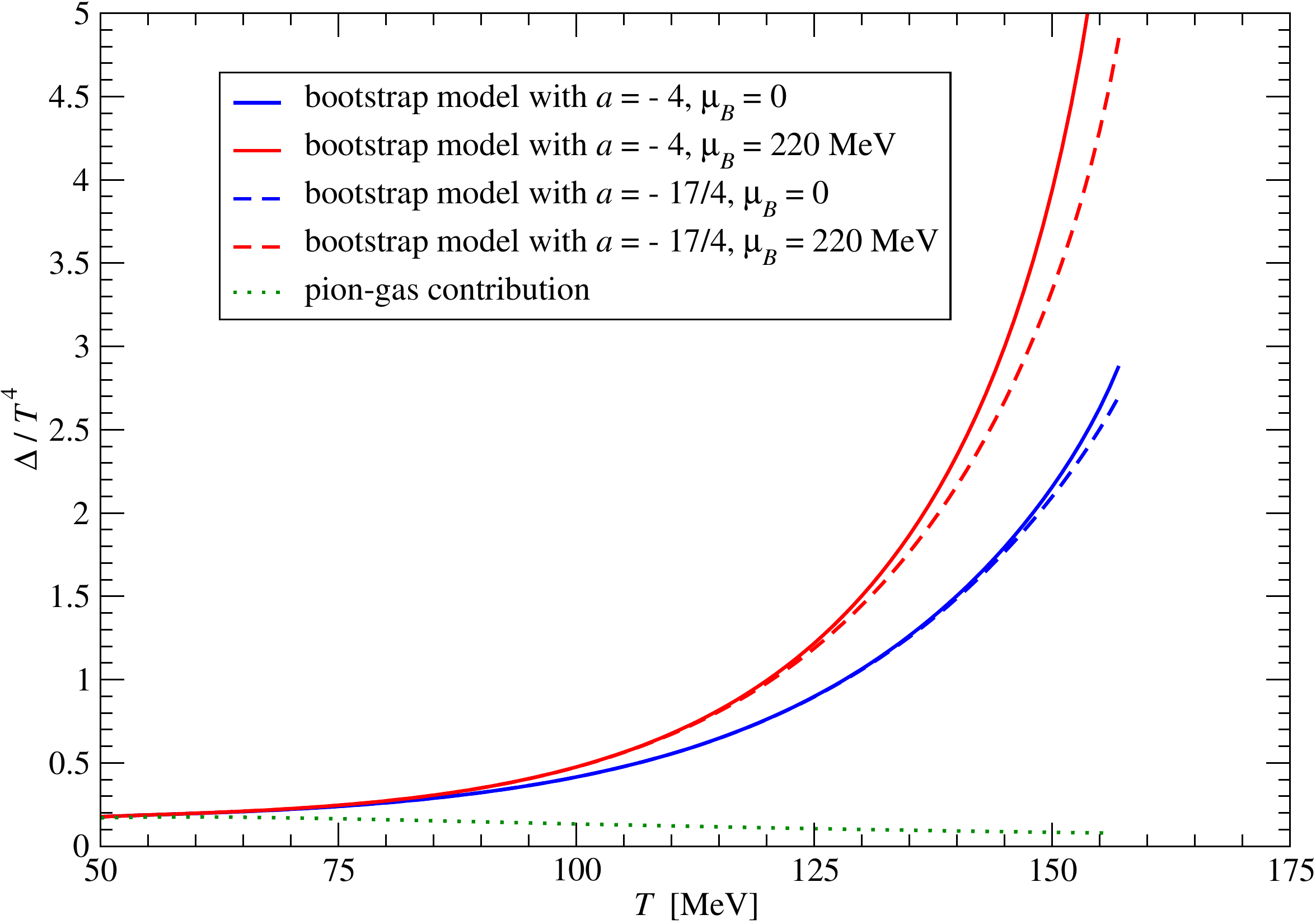} \hfill \includegraphics[width=0.48\textwidth]{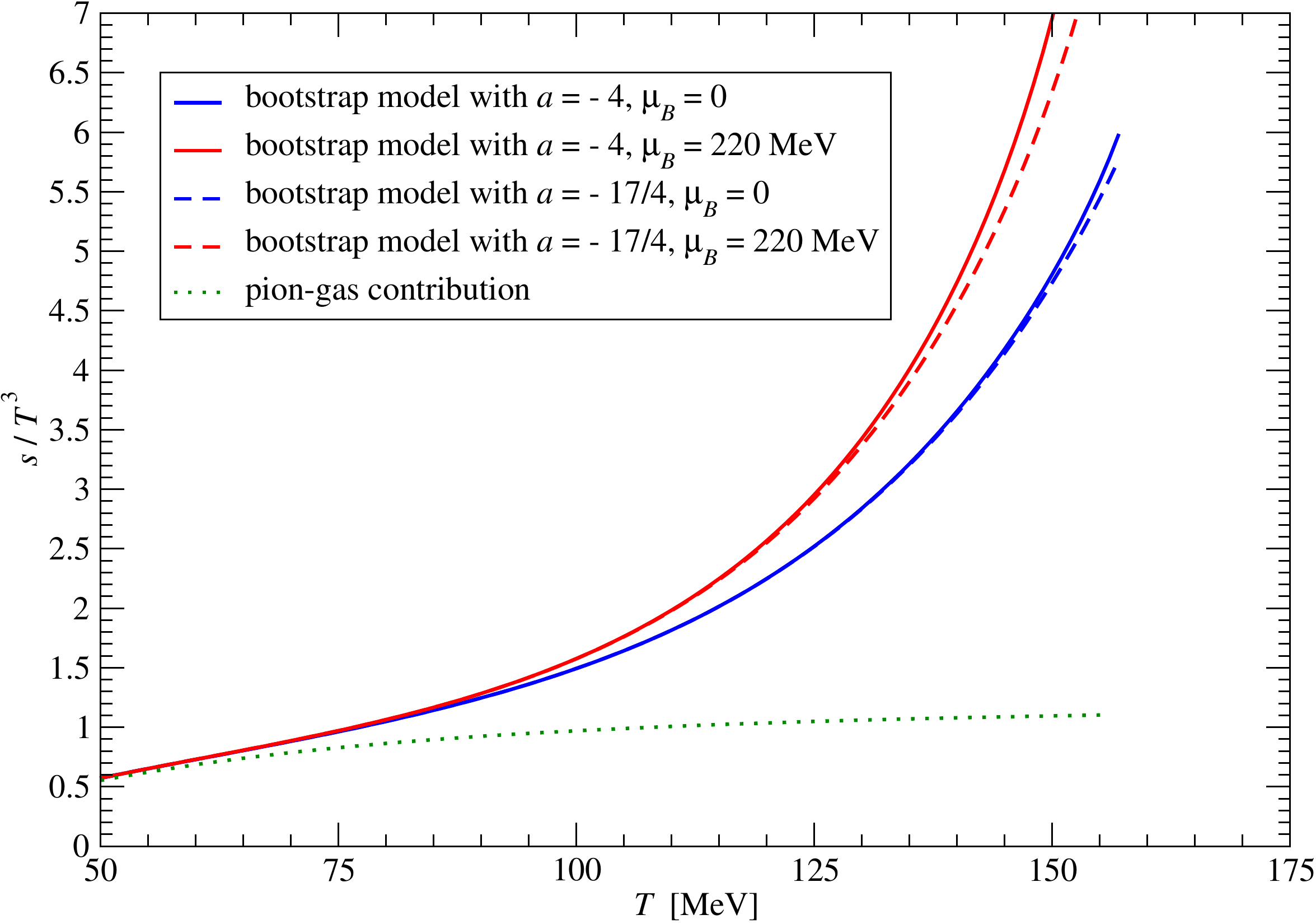}}
\caption{Equilibrium thermodynamics quantities for different types of continuous resonance spectrum distributions, as a function of the temperature $T$, in MeV. The four panels show the pressure $p$ (top left), the energy density $\varepsilon$ (top right), and the trace of the energy-momentum tensor $\Delta=\varepsilon-3p$ (bottom left) in units of $T^4$, and the entropy density $s$ (bottom right) in units of $T^3$. The solid curves correspond to $\rho_1(m)=A_1\:m^{a}\exp(bm)$ with $a=-4$, whereas the dashed curves are obtained for $a=-17/4$. The quark chemical potential is assumed to be $\mu_B=0$ for the blue curves, while the red curves are obtained at $\mu_B=220$~MeV. The parameter $b$ is set to the inverse of the critical Hagedorn temperature, as discussed in the paragraph after eq.~(\ref{hagspec}). In addition, we also plot the curves representing the contribution due to an ideal pion gas (dotted green curves), i.e. to the lightest states in the {\it{discrete}} part of the spectrum in eq.~(\ref{states}), which does not depend on the functional form that is assumed to model the {\it{continuous}} part of the spectrum.}
\label{thermo}
\end{center}
\end{figure}
Before discussing the behavior of viscosity coefficients near $T_c$, it is instructive to point out a few remarks about the thermodynamics of the model. In table~\ref{tbl_hag}, we report the parameters of the continuous part of the density of states, taken from refs.~\cite{Castorina:2009de, NoronhaHostler:2012ug}. Note that, as discussed in those references, the continuous part of the density of states is assumed to start at mass values corresponding to the pion-pair threshold, and that, in addition to the continuous part, the density of states also includes a $\delta$-like contribution at the pion mass. From the density of states constructed using the parameters in table~\ref{tbl_hag}, one obtains the equilibrium thermodynamic quantities shown in fig.~\ref{thermo}, namely the pressure ($p$), the energy density ($\varepsilon$) and the entropy density ($s$), in units of $T^4$ (for $p$ and $\varepsilon$) and $T^3$ (for $s$). We have also plotted the trace of the energy-momentum tensor $\Delta$ in units of the fourth power of the temperature, $\Delta/T^4=(\varepsilon-3p)/T^4$. The solid blue curves correspond to a continuous density of states of the form $\rho_1(m)=A_1\:m^{-4}\exp (b m)$ at vanishing chemical potential, whereas the solid red ones are obtained at $\mu_B=220$~MeV. To give an idea of the dependence of equilibrium thermodynamic quantities on $a$, we also show the results that one would obtain for a different value of $a$, i.e. for a spectral density with continuous part $\rho_1(m)=A_1\:m^{-17/4}\exp (b m)$, which are shown, with the same color code, by the dashed curves. One immediately realizes that, as compared with the solid curves, the dashed ones exhibit only small quantitative differences. The reason for the choice $a=-17/4$ stems from the fact that, as was discussed in subsection~\ref{subsec:critical_exponents_in_SBM}, the specific heat exhibits power-law behavior only if $a$ is larger than $-9/2$. On the other hand, we also remarked that $a$ is constrained to be less than $-7/2$, because in this range the energy density remains finite when $T$ tends to the critical temperature. This leaves us with $(-9/2,-7/2)$ as the most interesting interval of values for $a$. Thus, $a=-17/4$ is a value which is exactly equidistant from our choice $a=-4$ and the lower end of the interval of interesting values, and as such is expected to reveal some information on the dependence of our results on the choice of $a$. As the plots in fig.~\ref{thermo} clearly show, this dependence is very mild, indicating that our predictions for these quantities are robust (at least within the interval of $a$ values, i.e. $-9/2 \le a \le -7/2$).

Finally, the dotted green curves show the contributions from the ideal pion gas, i.e. the lightest hadrons included in the discrete and model-independent part of the density of states in eq.~(\ref{states}), which can be directly derived from eq.~(\ref{pion_contribution_to_ln_Z}): for example, the pion-gas contribution to the pressure (that one can denote as $p_\pi$) can be written as
\begin{equation}
\label{pion_contribution_to_pressure}
p_\pi=\frac{T}{V}\ln\mathcal{Z}_{\pi}=\frac{g_{\pi}}{2\pi^2}m_\pi^2 T^2 \sum_{n=1}^\infty \frac{K_{2}(n m_\pi /T)}{n^2}.
\end{equation}
It is known from comparison with lattice QCD results (as reviewed, for instance, in the recent ref.~\cite{Dexheimer:2020zzs}) that the hadron resonance gas model provides a very accurate description of the equation of state for all temperatures below $\Tc$. The contribution to thermodynamics from the part of the hadronic spectrum that is modelled in terms of a continuous density of states becomes significant when the temperature is sufficiently large. Nevertheless, in the case of $\rho_1$ with $a=-4$ both the energy and the entropy densities remain finite for $T\to \Tc^-$. This reflects the fact that, for $a=-4$, the second derivative of the partition function is divergent, but the first is not. In fact, setting $A_1=15\Tc^3=0.06144$~GeV$^3$ corresponds to $\varepsilon_c/\Tc^4\simeq 4$~\cite{Castorina:2009de}.

It is worth noting that the bootstrap model predicts the existence of a phase transition at the finite critical temperature $\Tc$. This can be interpreted by saying that this phenomenological model, which provides a description for the thermodynamics of hadronic matter in rather simple terms (e.g. neglecting hadron-hadron interactions) and without reference to the microscopic QCD Lagrangian, is able to capture the existence of a finite temperature, above which hadrons cannot exist anymore. To draw an analogy with the description of physics at the electro-weak scale within and beyond the Standard Model, the statistical bootstrap model can be interpreted as an ``effective field theory'' describing the thermal properties of nuclear matter in terms of its ``low-energy degrees of freedom'' (i.e. those that manifest themselves at energy scales below the characteristic hadronic scale, $O(10^2)$~MeV), and its breakdown at a finite temperature $\Tc$ hints at the existence of ``new physics'' above that scale. In this case, the ``new physics'' above that temperature is the quark-gluon plasma, whose existence could be argued (and reconciled with the bootstrap model~\cite{Hagedorn:1965st, Frautschi:1971ij}) after the introduction of QCD~\cite{Cabibbo:1975ig}. In this analogy, QCD plays the role of the ``more fundamental theory'', which holds up to higher energies (being, in fact, a renormalizable, asymptotically free and ultraviolet-complete theory) and at the same time reduces to the ``effective model'' at low energies, by predicting the existence of massive hadrons through the mechanisms of color confinement and dynamical chiral symmetry breaking~\cite{Kronfeld:2012uk}. One should remark that, despite the remarkable qualitative prediction of a finite maximal temperature at which hadrons exist, the bootstrap model does not capture all quantitative details of the change of phase between hadronic matter and the quark-gluon plasma: in particular, non-perturbative lattice calculations based on the QCD Lagrangian show that, for zero or nearly zero values of the baryonic chemical potential, this change of phase is actually an analytical crossover, rather than an actual phase transition (see refs.~\cite{Szabo:2014iqa, Ding:2015ona} and references therein). As a consequence, the statistical bootstrap model prediction (for $a=-4$ and at zero net baryon density) of a phase transition with critical exponents $\hat{\alpha}=1/2$, $\hat{\beta}=1$, and $\hat{\gamma}=0$ is disproven by lattice QCD. Still, the statistical bootstrap model remains a useful phenomenological model, in particular when studying regions of the phase diagram at large baryonic densities, where a critical endpoint might exist, and in which, as we already pointed out in section~\ref{sec:introduction}, lattice QCD calculations are hampered by particularly severe computational challenges.

\begin{figure}[h!]
\begin{center}
\centerline{\includegraphics[width=0.48\textwidth]{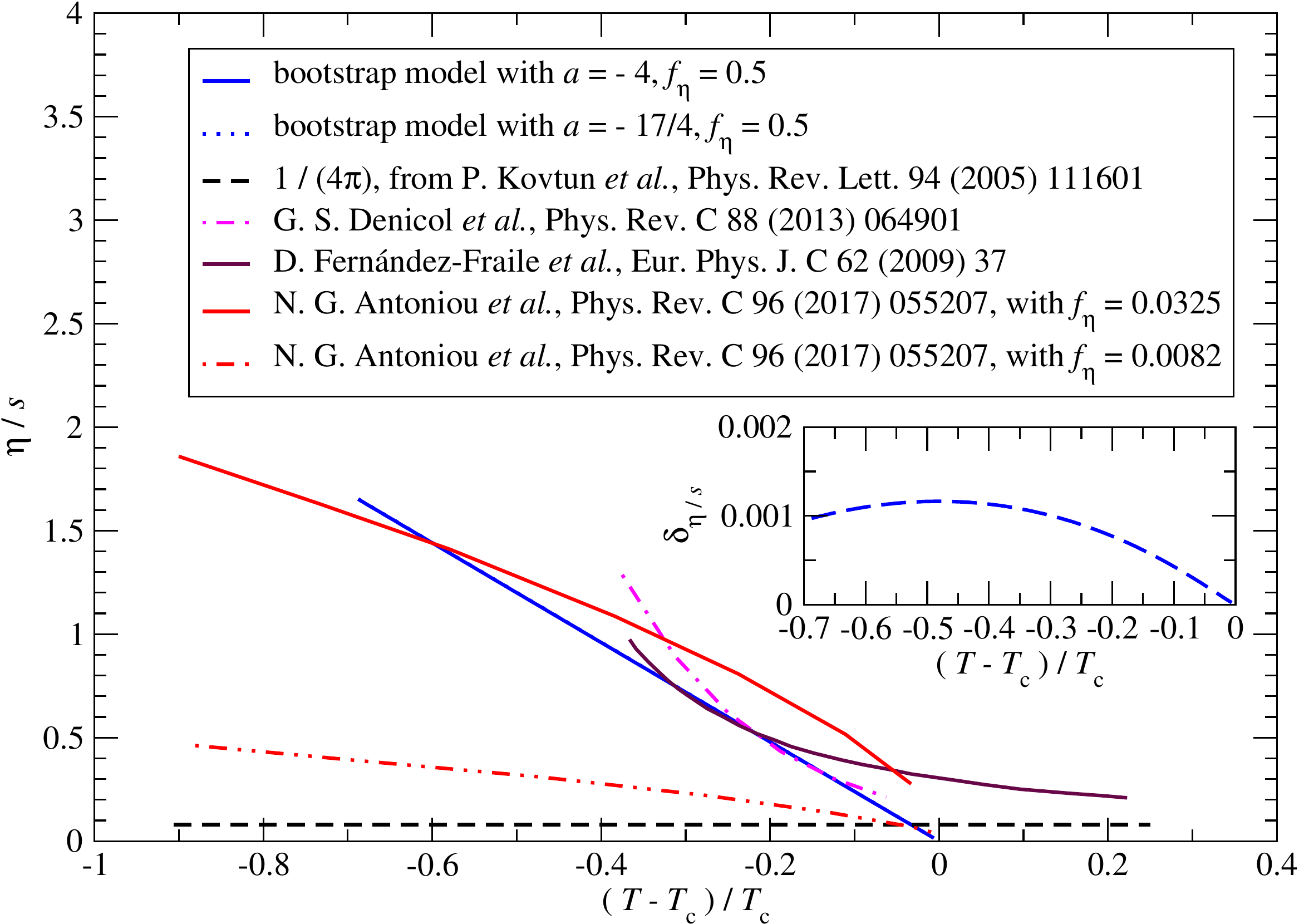} \hfill \includegraphics[width=0.48\textwidth]{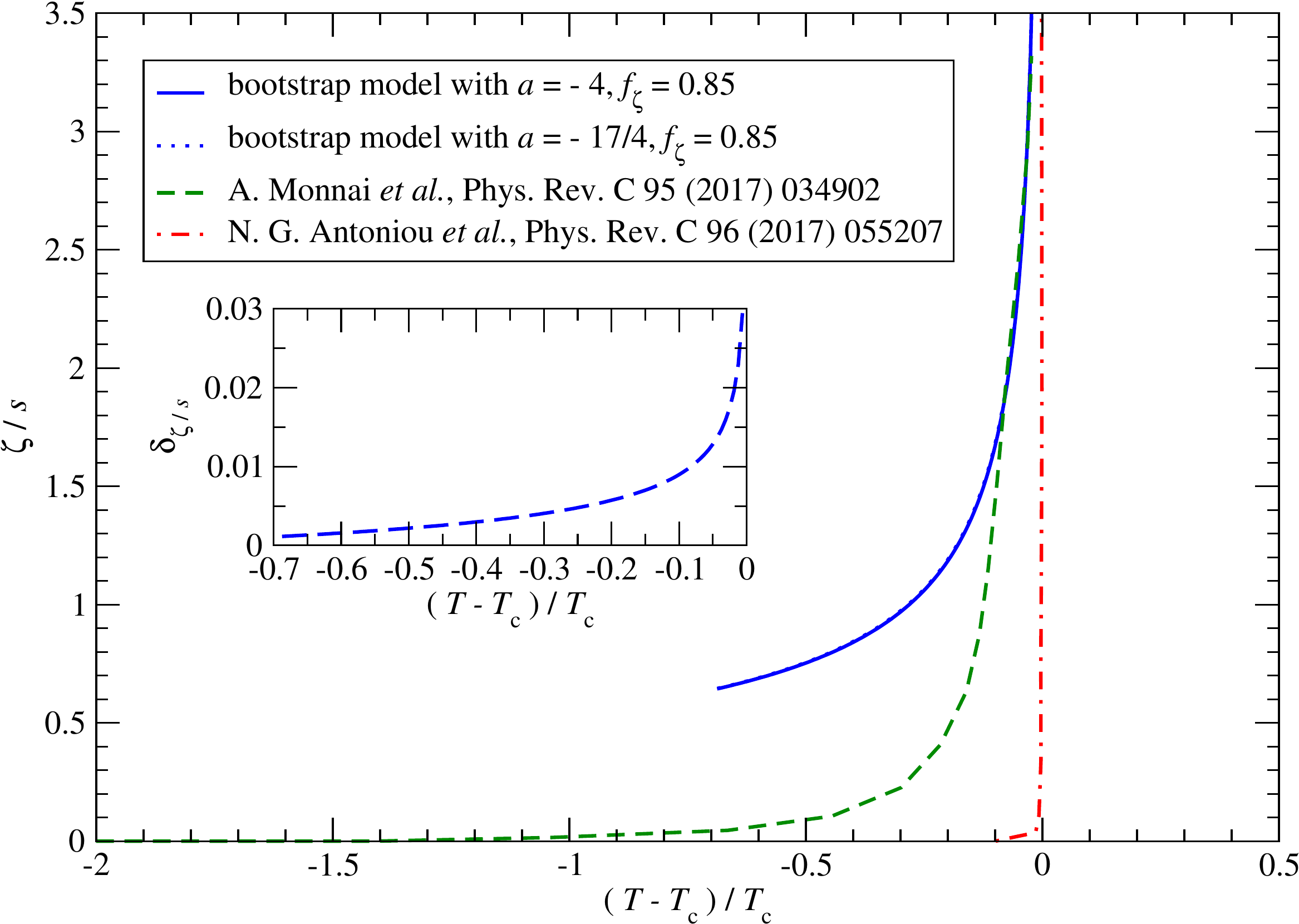}}
		\caption{The left-hand-side panel shows a comparison of the prediction of our model for the shear viscosity to entropy density ratio with those from other models~\cite{Antoniou:2016ikh,Kovtun:2004de,FernandezFraile:2009mi,Denicol:2013nua}. Our predictions for $a=-4$, denoted by the solid blue line, are also compared with those for $a=-17/4$ (shown by the dotted blue curve, which is nearly indistinguishable from the former), and the difference $\delta_{\eta/s}$ is plotted in the inset figure. The right-hand-side panel shows the prediction of the statistical bootstrap model for the bulk viscosity to entropy density ratio (for $a=-4$ and for $a=-17/4$, and the difference between the two, denoted by $\delta_{\zeta/s}$ and displayed in the inset figure) and its comparison with other works~\cite{Monnai:2016kud,Monnai:2017ber,Antoniou:2016ikh}. Our results correspond to $f_\eta=0.5$ and $f_{\zeta}=0.85$, which are fixed by requiring consistency with other models in the vicinity of the critical point, as discussed in the text.}
		\label{visc}
	\end{center}
\end{figure}

Fig.~\ref{visc} shows the predictions for the shear and bulk viscosities near the QCD critical point based on eqs.~(\ref{etatc}) and~(\ref{zetatc}). We take the correlation length amplitude to be $\Xi_{-}=1$~fm and the estimate for the critical point location to be $(\Tc,\mu_{B,\mbox{\tiny{c}}})=(160\mbox{~MeV}, 220\mbox{~MeV})$~\cite{Antoniou:2008vv}. For $a=-4$ one can easily derive the critical exponents and amplitudes needed for the estimate of the viscosity coefficients near $\Tc$. The critical exponents are not independent but are constrained by scaling laws. In particular, the exponents $\hat{\alpha}$ and $\hat{\nu}$ are related by the Josephson scaling law $\hat{\nu} d=2-\hat{\alpha}$, where $d$ is the number of space dimensions~\cite{Huang:1987sm}.

In the left-hand-side panel of fig.~\ref{visc}, the solid blue curve shows the shear viscosity to entropy density ratio within the statistical bootstrap model, with density of states specified by $\rho_1$, and with the $f_{\eta}$ parameter appearing on the right-hand side of eq.~(\ref{etatc}) set to $0.5$. Note that the choice of this amplitude value, which we have done with the procedure discussed below, introduces some systematic uncertainties. On the other hand, to give an idea of the dependence of this prediction on the parameter $a$, we also present the prediction that one would obtain for $a=-17/4$ (again with $f_{\eta}=0.5$), which is displayed by the dotted blue curve, and which is nearly indistinguishable from the latter. Hence, in the inset plot we show the quantity $\delta_{\eta/s}$, defined as the prediction of the bootstrap model for $\eta/s$ for $a=-17/4$, minus the one for $a=-4$: the relative difference between the predictions corresponding to the two $a$ values is at the per~mille level. We conclude that the dependence of our prediction on $a$ (within the range of values of $a$ of our interest) has a negligible impact on the uncertainties affecting the prediction for the $\eta/s$ ratio. In the larger plot in the figure, we also compare the critical solution for $\eta/s$ with those obtained from various other models: in particular, the dashed black curve corresponds to the conjectured universal lower bound $1/(4\pi)$ for this ratio, that was derived in ref.~\cite{Kovtun:2004de}, while the brown curve describes the result that one would obtain for a pion gas~\cite{FernandezFraile:2009mi}, and the magenta curve shows the result that can be derived assuming the medium to be described in terms of a hadronic mixture~\cite{Denicol:2013nua} at low temperature and density. Finally, the red curve corresponds to the same solution for the viscosity coefficients as in this work, but with the critical exponents of the three-dimensional Ising model and the amplitudes constrained by universality arguments~\cite{Antoniou:2016ikh}. In our case, we chose to fix the $f_{\eta}$ amplitude to optimize the consistency with the other predictions shown in the figure at temperatures $-0.2 \lesssim t \lesssim -0.1$: in particular, in that temperature interval our choice yields an almost perfect consistency with the curve predicted in ref.~\cite{Denicol:2013nua}, which is the one that is intermediate among those predicted in those works. We should remark, however, that in general the choice of the $f_{\eta}$ amplitude remains a source of systematics that are difficult to quantify (and, hence, the value that we quote should be taken \emph{cum grano salis}). Note, however, that, as shown by the two curves derived in ref.~\cite{Antoniou:2016ikh} with two different choices for $f_{\eta}$, i.e. the solid and dash-dot-dotted red lines, the choice of the numerical value of the amplitude has a strong impact at temperatures far from the critical point, but this discrepancy is already reduced to small values for reduced temperatures between approximately $-0.2$ and $-0.1$. We note that the critical behavior of the statistical bootstrap model leads to a linear decrease in $\eta/s$ as a function of the temperature, and that at low temperatures the estimated magnitude of $\eta/s$ is in agreement with that of a pion gas, or of the hadron gas mixture. Near $\Tc$ there is a mild violation of the bound conjectured in ref.~\cite{Kovtun:2004de} (which could make it problematic to fix $f_{\eta}$ through some constraint in a region of temperatures very close to $\Tc$). Such violation has also been noted for one of the solutions discussed in ref.~\cite{Antoniou:2016ikh}, shown by the dashed red curve in fig.~\ref{visc}.
 
The right-hand-side panel of fig.~\ref{visc} shows the bulk viscosity to entropy density ratio, in which one notes that the statistical bootstrap model predicts a rapid increase in the bulk viscosity as a function of the temperature. Also in this case, we present our results both for $a=-4$ (solid blue line) and for $a=-17/4$ (dotted blue curve), and the difference between the latter and the former, which is denoted by $\delta_{\zeta/s}$ and shown by the dashed blue line in the inset plot. In this case, the relative difference between the predictions corresponding to the two $a$ values is below $10^{-2}$, meaning that also for $\zeta/s$ the dependence on $a$ induces a very mild systematic uncertainty. Near $\Tc$, our results, with the amplitude coefficient appearing in eq.~(\ref{zetatc}) fixed to $f_{\zeta}=0.85$ by requiring an approximate match with those of refs.~\cite{Monnai:2016kud,Monnai:2017ber} at $t \simeq -0.1$, are in remarkable agreement with those from that work (shown by the dashed green curve), where the bulk viscosity has been estimated under the assumption of a QCD critical point belonging to the dynamical universality class of the so-called H model~\cite{Hohenberg:1977ym}. Remarkably, this agreement between the two curves is observed for essentially all negative values $t \gtrsim -0.1$, which is non-trivial, as that is the region in which the $\zeta/s$ ratio grows rapidly to very large values. Nevertheless, also in this case the readers should be warned that there is no obvious method to fix the value of $f_{\zeta}$ in a unique, completely rigorous way from first principles, and the systematic uncertainties associated with any choice remain difficult to assess. For comparison, in the plot we also show the prediction for $\zeta/s$ from ref.~\cite{Antoniou:2016ikh}. Coming to the interpretation of fig.~\ref{visc}, we note that a large bulk viscosity should manifest itself in heavy-ion collisions through the decrease of the average transverse momentum of  final-state hadrons. Moreover, due to the increase in entropy associated with the dissipation through large bulk viscosity, this effect should be accompanied by an increase in total multiplicity for final-state hadrons. The large bulk viscosity near the critical point would play a particularly important role in the elliptic flow measurement  of the matter produced in the BES program.

Note that the features of the transport coefficients predicted by our model are only expected to hold close to $\Tc$, and there is no reason to expect the curves plotted in fig.~\ref{visc} to be quantitatively accurate predictions at temperatures much smaller than the critical one. The reason for this was already discussed in ref.~\cite{Antoniou:2016ikh}, in which it was remarked that the extrapolation of power-law behavior beyond the critical region can be, at best, a crude approximation. Indeed, by definition, the critical exponents only capture the ``universal'' critical features of the system, not its full dynamics. Nevertheless, it is interesting to plot these quantities in a range of temperatures similar to the one that was used for the equilibrium thermodynamics quantities (for which, as we pointed out above, the predictions of our model are instead expected to extend to all temperatures below $\Tc$), which allows one to highlight, in particular, the monotonically decreasing dependence of $\eta/s$ as a function of the temperature for $T\le \Tc$, and the dramatic increase of $\zeta/s$ close to the critical point.

\section{Discussion and conclusions}
\label{sec:discussion_and_conclusions}

\subsection{Discussion}\label{subsec:discussion}

In this work we derived the predictions of the statistical bootstrap model for thermodynamic quantities and transport coefficients near the critical endpoint of QCD. While it is well known that equilibrium thermodynamic quantities at temperatures below the (pseudo-)critical one are described well in terms of a gas of non-interacting hadrons, when all experimentally observed hadronic states with masses up to approximately $2$~GeV~\cite{Vovchenko:2014pka, HotQCD:2014kol} are included, the introduction of a continuous, Hagedorn-like, density of states for heavier states in the spectrum leads to the manifestation of critical behavior, without substantially altering the predictions for the equation of state at low temperatures. Moreover, as we remarked, the phenomenological implications of the model do not depend on the precise value of $M$, which in our computation was set to $2.25$~GeV.

Even though the derivation of the critical exponents for this model is based on the assumption of a spectral density valid at zero chemical potential, and the dependence on $\mu_B$ of the logarithm of the partition function is simply encoded in a fugacity factor, we argued that it may still capture the correct physics close to a possible critical point at finite $\mu_B$. For a continuous density of states of the form $\rho(m)\sim m^{a}\:\exp \left (bm \right)$ with $a<-7/2$, the energy and entropy densities remain finite even for point-like hadrons, while all higher-order derivatives diverge near $\Tc$. For $a=-4$ the energy density remains finite as $T\to \Tc$, signalling the existence of a phase transition. 

In passing, it is worth mentioning that a continuous spectral density of the form required for self-consistency in the statistical bootstrap model, eq.~(\ref{hagspec}), also arises if one models hadrons in terms of confining strings of glue (as was done, for example, in ref.~\cite{Isgur:1984bm}).

Next, we studied the critical behavior of shear and bulk viscosities within the statistical bootstrap model. Identifying the thermodynamic quantities whose singular parts would contribute to the viscosity coefficients it is possible to write down \emph{Ans\"atze} for the viscosity coefficients valid near the critical point. Using the \emph{Ans\"atze} in eqs.~(\ref{eta_visc_Tc}) and~(\ref{zeta_visc_Tc}) together with the singular part of the relevant thermodynamic quantities in eqs.~(\ref{alpha_exponent})--(\ref{nu_exponent}), one can obtain the dominating contributions for the viscosity coefficients in eqs.~(\ref{eta_visc_Tc}) and~(\ref{zeta_visc_Tc}). We found that the statistical bootstrap model predicts the shear viscosity to entropy density ratio $\eta/s$ to decrease quite rapidly near $\Tc$. We observe that the magnitude of $\eta/s$ away from the critical point is in good agreement with the predictions of non-critical models, and that there is a mild violation of the $\eta/s\ge 1/(4\pi)$ bound~\cite{Kovtun:2004de} near $\Tc$. It is worth emphasizing that this (slight) violation of the $\eta/s\ge 1/(4\pi)$ bound might be unphysical, i.e. an artifact of the model. In fact, it is also worth remarking that, while the conjecture of the $\eta/s\ge 1/(4\pi)$ bound was first derived in a holographic context~\cite{Kovtun:2004de} (and is expected to be saturated in strongly coupled gauge theories with a known gravity dual~\cite{Buchel:2003tz}, such as the $\mathcal{N}=4$ supersymmetric Yang-Mills theory~\cite{Policastro:2001yc}), its origin is, in fact, much more general, being related to the uncertainty principle of quantum systems. As for the $\zeta/s$ ratio, we found it to rise very rapidly near $\Tc$, in remarkable agreement with refs.~\cite{Monnai:2016kud,Monnai:2017ber}.  

The anomalous behavior of shear and bulk viscosity coefficients near the critical endpoint might be very important for heavy-ion collision experiments. In particular, an enhanced bulk viscosity should manifest itself in heavy ion collisions through a decrease of the average transverse momentum of  final state hadrons, and a corresponding  increase in entropy. This feature may be particularly important for the experimental search for the critical endpoint of QCD through the BES program.

It is interesting to discuss a comparison of our findings with other related works. In particular, a study similar to ours was recently reported in ref.~\cite{Rais:2019chb}, which also predicts the transport properties of hot QCD matter within a hadron resonance model, finding a very low shear viscosity to entropy density ratio near $\THag$, in agreement with our results. In contrast to our work, however, the focus of that article is not on the behavior near criticality (where, as we have discussed in detail above, much information can be derived with purely analytical calculations and general universality arguments), but instead on their numerical determination in a wider range, using the Gie{\ss}en Boltzmann-Uehling-Uhlenbeck transport model~\cite{Buss:2011mx} and Monte~Carlo calculations. The fact that the numerical approach used in ref.~\cite{Rais:2019chb} reproduces our analytical results close to criticality is a non-trivial cross-check of the results.

We should emphasize again that there is no fundamental proof that the statistical bootstrap model described in this work should necessarily provide a complete description of the fundamental properties at the CEP. Based on very general arguments (including the continuous nature of the transition at the critical endpoint, spacetime dimensionality, and the underlying symmetries---or lack thereof---of the theory), one may instead argue that the universality class of the critical endpoint of QCD should instead be the one of the Ising model in three dimensions (for a review, see ref.~\cite{Pelissetto:2000ek}). The critical exponents in this model have recently been computed to very high precisions using conformal bootstrap techniques~\cite{ElShowk:2012ht, El-Showk:2014dwa}, finding~$\hat{\alpha}=0.11008(1)$, $\hat{\gamma}=1.237075(10)$, etc., which are clearly incompatible with those predicted by the model that we considered here. If the critical endpoint of QCD exists, it may well be that its actual critical exponents are those of the three-dimensional Ising model, and that deviations from the description in terms of the statistical bootstrap model start to occur when one approaches the CEP. In this respect, it would be interesting to study theoretically how these deviations start to manifest themselves when the system is off, but close to, the critical point---perhaps using the analytical tools of conformal perturbation theory~\cite{Guida:1995kc} (see also ref.~\cite{Caselle:2016mww}, for an explicit example of application), as recently proposed in ref.~\cite{Caselle:2019tiv}. An interesting issue associated with the description of the critical endpoint of QCD in terms of the three-dimensional Ising model concerns the identification of the lines, in the QCD phase diagram, that describe relevant deformations of the model (see also ref.~\cite{Caselle:2020tjz}): what are the directions that correspond to a ``thermal'' and to a ``magnetic'' perturbation of the critical point? How do they affect the reliability of the description of the thermodynamics of the hadronic phase in terms of a hadron resonance gas with a spectrum of the form in eq.~(\ref{states})? The answers to these questions may have important phenomenological implications for the evolution of the medium in energy scans going through or close to the critical endpoint, since they could directly affect the dynamics of hadrons before freeze-out.

Finally, it should be noted that, by construction, the statistical bootstrap model does not allow one to describe the approach to the critical endpoint of QCD from ``above'', i.e. from the deconfined phase. Perturbative computations show that the $\eta/s$ ratio is generally large for a weakly coupled quark-gluon plasma~\cite{Hosoya:1983xm, Thoma:1991em} (a seemingly counter-intuitive result, which, in fact, reflects the fact that suppression of interactions makes ``transverse'' propagation of momentum difficult), but it is well known that thermal weak-coupling expansions are affected by non-trivial divergences, which are not present at $T=0$ (see, for example, ref.~\cite{Ghiglieri:2020dpq} and references therein), and require a sophisticated treatment~\cite{Jeon:1995zm, Arnold:2000dr, Arnold:2003zc, Arnold:2002zm}.

\subsection{Conclusions}
\label{subsec:conclusions}

To summarize, in this work we derived the theoretical predictions of the statistical bootstrap model in the vicinity of the critical endpoint of QCD. Working in the approximation in which the critical temperature does not depend on the value of the chemical potential (which, as we remarked in section~\ref{subsec:formulation_of_the_model}, has support from lattice QCD calculations showing that the crossover line in the phase diagram of the theory has very small curvature~\cite{Kaczmarek:2011zz, Endrodi:2011gv, Cea:2014xva, Bonati:2014rfa}), we showed that, for a suitable choice of its parameters, this model ``predicts'' the existence of a phase transition, and allows one to derive the associated critical exponents in an analytical way. Moreover, the model also gives predictions for the transport properties near the CEP, which are encoded in the shear and bulk viscosities. Both for the equation of state and for these transport coefficients, the dependence of the predictions of the model on the parameter $a$ (within the rather narrow interval of physical interest) is very mild. In spite of the relative simplicity of the model, these results are qualitatively and quantitatively very similar to those obtained from other approaches. These findings may hopefully guide the future experimental identification of the CEP in heavy-ion collision experiments and the determination of the physical properties of QCD matter in the proximity of the critical endpoint.

\vspace{5mm}
\section*{Acknowledgments}
Guruprasad~Kadam is financially supported by the DST-INSPIRE faculty award under Grant No. DST/INSPIRE/04/2017/002293. G.K. thanks Sangyong~Jeon for useful comments.

\bibliography{paper_tcv1}

\end{document}